\documentclass[prl,aps,floatfix,showpacs,tightenlines,twocolumn,superscriptaddress,amsmath,amssymb]{revtex4}

\usepackage{graphicx}% Include figure files
\usepackage{dcolumn}% Align table columns on decimal point
\usepackage{bm}% bold math
\usepackage{epsfig}

\def\comment#1{}

\begin{document}

\title{A high bandwidth quantum repeater}

\author{W. J. Munro}\email{bill.munro@hp.com}
\affiliation{Hewlett-Packard Laboratories, Filton Road, Stoke Gifford, Bristol BS34 8QZ, United Kingdom} 
\affiliation{National Institute of Informatics, 2-1-2 Hitotsubashi, Chiyoda-ku, Tokyo 101-8430, Japan}

\author{R. Van Meter}
\affiliation{National Institute of Informatics, 2-1-2 Hitotsubashi, Chiyoda-ku, Tokyo 101-8430, Japan}
\affiliation{Keio University, 5322 Endo, Fujisawa, Kanagawa, 252-8520, Japan}

\author{Sebastien G.R. Louis}
\affiliation{National Institute of Informatics, 2-1-2 Hitotsubashi, Chiyoda-ku, Tokyo 101-8430, Japan}
\affiliation{Department of Informatics, School of Multidisciplinary Sciences, The Graduate University for 
Advanced Studies, 2-1-2 Hitotsubashi, Chiyoda-ku, Tokyo 101-8430 Japan}

\author{Kae Nemoto}
\affiliation{National Institute of Informatics, 2-1-2 Hitotsubashi, Chiyoda-ku, Tokyo 101-8430, Japan}

\begin{abstract}
We present a physical- and link-level design for the creation of
entangled pairs to be used in quantum repeater applications where one
can control the noise level of the initially distributed pairs.  The
system can tune dynamically, trading initial fidelity for success
probability, from high fidelity pairs (F=0.98 or above) to moderate
fidelity pairs.  The same physical resources that create the
long-distance entanglement are used to implement the local gates
required for entanglement purification and swapping, creating a
homogeneous repeater architecture. Optimizing the noise properties of
the initially distributed pairs significantly improves the rate of
generating long-distance Bell pairs. Finally, we discuss the
performance trade-off between spatial and temporal resources.
\end{abstract}


\maketitle

Quantum information has reached a very interesting stage in its
development, where we have seen many fundamental experiments laying
the foundation for practical systems\cite{spiller05}.  Now certain
applications, such as quantum key distribution (QKD), are being
readied for commercial use\cite{gisin02,Duligall06}, where practical
distances hover around the 150km mark.  Any quantum communication
longer than this limit suffers severely from noise and exponential
loss in the quantum communication channel.  Hence, the quantum
communication for either QKD over a distance beyond the limit or, more
generally, all distributed quantum information processing, requires
the development of a high-bandwidth repeater which can distribute and
potentially process quantum information given these constraints.  In a
quantum repeater system, initial imperfect Bell pairs (which we call
base-level pairs) are distributed over channel segments. These base
pairs are then purified to high fidelity Bell pairs and connected via
entanglement swapping, resulting in entanglement between the qubits at
distant stations. Iterating this procedure creates
Bell pairs at even larger distances.  These pairs can be used in many
different applications, including QKD, quantum communication,
distributed quantum computation, and quantum metrology and related
uses\cite{van-meter07,cirac99,grover97,serafini06,chuang00}.

The many recently-proposed schemes for the design of a quantum
repeater fall into two categories. The majority of the schemes focus
on the heralded creation of very high fidelity base-level
pairs\cite{briegel98,dur98,childress06,klein06,enk98,
  duan01,chen07,chen07a,jiang07}. For longer segment lengths, the
generation of these high fidelity pairs comes at the expense of a very
low probability of success, which becomes one of the major bottlenecks
in the overall performance of a repeater system.  Another significant
issue is that in the majority of these schemes, local gates between
multiple qubits within a single repeater station are difficult. An
alternative approach has recently been proposed, instead creating
base-level pairs of moderate fidelity and high heralded success
probability\cite{loock06,ladd06}.  In this second approach,
the physical resources used for long-distance entanglement also
efficiently implement local gates, facilitating the purification of
moderate fidelity pairs back to high fidelity pairs.  For instance with 
16 qubits/node with a 10km spacing rate 15 pairs of fidelity  F=0.98 can 
be achieved over a 1280km repeater network, however at longer repeater node 
spacing distanes ($>40$km) the rate fails to zero. This node spacing isssue is 
one of the key limitations for this second approach. 
These two approaches are radically different in the use of physical resources
and in technological requirements.  It is hence not trivial to
directly compare the feasibility and efficiency of schemes in
different approaches.  However, it has been thought that these two
approaches are complementary to each other, trading high fidelity for
high success probability or vice versa.  Quantum repeaters of the
latter kind typically use coherent light instead of the single photons
common in the former category.  It has been believed that a
coherent-light quantum repeater is fundamentally unable to generate
high-fidelity Bell pairs and hence is unable to cope with severe loss
in the quantum channel. In this letter we address this shortcoming by
presenting the design of a new scheme for entanglement distribution
utilizing coherent light and demonstrate that in fact such a system is
more flexible over a wide range of losses without serious overhead in
physical resources. This advance will have a significant impact on the
overall repeater performance.
%  hence the design of the  base-level mechanism for creating 
%  entanglement pairs carries fundamental XXXX impact on it.

In this letter we consider the design of a repeater segment where one
can dynamically vary the quality of the base-level entangled pairs
from very high to moderate fidelity.  This tuneability will allow one
to trade the fidelity of the entangled pairs against the probability
of their successful distribution for a given segment length. We will
also ensure that the same physical resources can be used to implement
the local gates necessary for the entanglement purification and
entanglement swapping\cite{nemoto04,spiller06}. In this case, both
requirements can be met via a controllable interaction between our
qubit and light field.

The core interaction between our qubit and field in our cavity quantum
electrodynamics (CQED) system is the Jaynes-Cummings Hamiltonian given
by $H_{JC} = \hbar g \left( a^\dagger \sigma_-  + a\sigma_+ \right)$ where 
$2g$ is the vacuum Rabi splitting for the dipole transition. $a$ ($a^{\dagger}$) 
refers to the annihilation (creation) operators of the electromagnetic
field mode in a cavity and $\sigma_+$ ($\sigma_-$) the raising (lower)
operators of the qubit with ground state $|0\rangle$ and excited state
$|1\rangle$. Our qubit is a solid-state electronic spin 
which is also coupled to a nuclear spin qubit,
allowing for a coherent transfer of quantum information to the
long-lived nuclear spin qubit. Physically, the electronic and
nuclear-spin systems may be achieved, for example, by single electrons
trapped in quantum dots, NV centers in diamond or neutral donor
impurities in semiconductors.  

Our basic J-C Hamiltonian can be used to implement a controlled
displacement $D\left(\beta \sigma_z\right)\equiv e^{\sigma_z
  \left(\beta a^\dagger-\beta^* a\right)}$
operation\cite{spiller06,loock07} between the qubit and field, where
$\sigma_z=|0\rangle \langle 0| - |1\rangle \langle 1|$. This operation
displaces the field mode by an amount $\beta$ conditioned on the state
of the qubit. There are a number of ways this interaction can be
achieved ranging from shaped pulse sequences modulating the qubit or
field\cite{milburn} to sequences of controlled rotations and
unconditional displacement operations\cite{loock07,footnote}. 
% The controlled rotations operations are achieved by operating in the 
% dispersive limit while unconditional displacements are achieved by 
% pumping of the probe field.

% This basic controlled rotation interaction can be used to implement a much more powerful operation, 
% the controlled displacement $D\left(i \beta \sigma_z\right)$\cite{spiller06,loock07},  by combining controlled 
% rotations with unconditional displacements, $D(\alpha\cos\theta)e^{-i \theta \sigma_z a^{\dagger} a} 
% D(-2\alpha)e^{i \theta \sigma_z a^{\dagger}a}D(\alpha\cos\theta)=D\left(i 2 |\alpha| \,\sin\theta  \sigma_z\right) 
% \equiv D\left(i \beta \sigma_z\right)$. Such an operation displaces the probe field by an amount 
% $\pm i \beta$ depending on the state of the qubit. Here $D\left(\gamma \right)=e^{\gamma X(\phi)}$ 
% where  $X(\phi)=(a^{\dagger} e^{i \phi} + a e^{-i\phi})$ is the field quadrature operator. We have 
% described a sequence of operations to implement the controlled displacement above; however, these can 
% be combined into a single operation  using an appropriately shaped pulse\cite{munro07}.

\begin{figure}[!htb]
\begin{center}\includegraphics[scale=0.45]{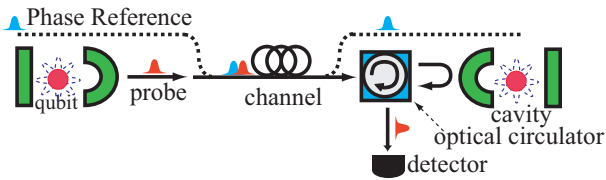}\end{center}
\caption{Schematic of an entanglement distribution scheme based on two qubits in individual cavities interacting indirectly 
via a shared probe beam and controlled displacement operations. An optical circulator before the second qubit routes the probe 
field into the cavity and then the probe beam leaking out of the cavity to the detector. A phase reference is sent along 
the same lossy channel.}
\label{figexp}
\end{figure}

These controlled displacement operations now form the basis of an efficient and tunable entanglement distribution 
scheme, as depicted in Fig (\ref{figexp}). The scheme works as follows: our first qubit is prepared in an equal 
superposition of both basis states $(|0\rangle+|1\rangle)/\sqrt{2}$ with the probe mode in the cavity initially 
being the vacuum. The qubit interacts with the probe beam via the controlled displacement operation 
$D\left(\beta \sigma_z\right)$  resulting in the combined qubit-probe state 
$(|0\rangle |\beta\rangle+|1\rangle |-\beta\rangle)/\sqrt{2}$. The probe beam is either switched 
out or leaks out of the cavity and is transmitted over the noisy loss channel to the second cavity. 
Here the second qubit and probe mode interact via a controlled displacement operation. The probe beam then 
leaks out of this second cavity and is measured, projecting our qubits into an appropriate entangled state. 
In more detail, if our two distributed qubits are prepared in a state $(|00\rangle+|01\rangle+
|10\rangle+|11\rangle)/ 2 $ then after both controlled displacements and the noisy channel our two 
qubit-light field state can be represented by
\begin{eqnarray}\label{systemstate}
\rho=\frac{1+e^{-\bar \gamma/2}}{2}|Z_+\rangle \langle Z_+| +\frac{1-e^{-\bar \gamma/2}}{2} |Z_-\rangle \langle Z_-|
\end{eqnarray}
where $|Z_\pm\rangle =\frac{\sqrt{\cal{N_\pm}}}{2} |\Phi_+\rangle |c_\pm\rangle+\frac{\sqrt{\cal{N_\mp}}}{2}|\Phi_-\rangle|c_\mp\rangle+
\frac{1}{\sqrt{2}} |\Psi_\pm\rangle |0 \rangle$, $|\Phi_\pm\rangle=|\frac{00\pm11}{\sqrt{2}}\rangle$, $|\Psi_\pm\rangle=|\frac{01\pm10}{\sqrt{2}}\rangle$, 
$|c_\pm\rangle =( |2 \beta e^{-l/2l_0} \rangle\pm|-2 \beta e^{-l/2l_0} \rangle)/ \sqrt{2\cal{N_\pm}}$ with 
${\cal N}_\pm=1\pm \exp \left[-8 |\beta|^2 e^{-l/l_0}\right]$ and  $\bar \gamma=2 |\beta|^2 (1-e^{-l/l_0})$. 
Here $l/l_0$ represents the attenuation of the probe in the channel. 	       
Now the probe beam in Eqn (\ref{systemstate}) is in one of three possible states, $|c_\pm\rangle$ or the vacuum $|0\rangle$.
The odd cat state $|c_-\rangle$ is orthogonal to both $|c_+\rangle$ and $|0\rangle$. However, $|c_+\rangle$ and $|0\rangle$ 
are non-orthogonal with an overlap $2 \exp \left[-4 |\beta|^2 e^{-l/l_0} \right]/(1+\exp \left[-8 |\beta|^2 e^{-l/l_0}\right])$. 
Of course, if $\beta e^{-l/2l_0} \gg 1$, these two states are effectively orthogonal, and so one could distinguish all 
probe beam states.  However, the channel loss has had two significant effects: first, it has mixed $|Z_+\rangle$ with 
$|Z_-\rangle$  with a mixing parameter $\frac{1+e^{-\bar \gamma/2}}{2}$. This mixing parameter is small only when 
$|\beta|^2 (1-e^{-l/l_0}) \ll 1$, which is in conflict with the desire
to have all probe beam states nearly orthogonal for channels of moderate 
length.  Second, the channel has also attenuated the probe beam's amplitude and so the second controlled displacement must be by 
$D\left(\beta e^{-l/2l_0} \sigma_{z_2}\right)$ rather than $D\left(\beta \sigma_{z_2}\right)$ to minimize the effect of the loss. 
% This reduction in the amplitude can be achieved by also transmitting the pump beam along the same noisy channel as the probe beam. 
%The pump beam is automatically attenuated by the required amount and so auto-calibrates the size of the unconditional displacement. 
% It also maintains phase stabilization between the probe and pump.

We now turn our attention to the measurement of the probe and the
resulting conditioning it causes on the distributed qubits.  There are
various measurement strategies, ranging from highly idealized cat
state projectors (CSP) $|c_-\rangle \langle c_- |$, to single photon
detection (SPD). Using such detection strategies, our qubits are
conditioned to
\begin{eqnarray}\label{rho_c}
\rho (F)=F|\Phi_-\rangle \langle \Phi_- | +(1-F) |\Phi_+\rangle \langle \Phi_+|
\end{eqnarray}
with heralded success probabilities
\begin{eqnarray}\label{prob1}
P_{CSP}[F,l/l_0]&=&\frac{1}{4} \left\{1- \left[2 F-1\right]^\frac{8 e^{-l/l_0}}{1-e^{-l/l_0}} \right\} \\
\label{prob2}
P_{SPD}[F,l/l_0,\eta^2]&=& \frac{1}{2}\left. \frac{d}{d\lambda}\left[ 2F-1\right]^{\frac{4\eta^{2}e^{-l/l_0}}{1+\left( 7- 8\eta^{2}\right) e^{-l/l_0}}\lambda}\right\vert _{\lambda=1}
\end{eqnarray}
respectively. The single photon detector is assumed to have a non-unit quantum 
efficiency $\eta^2$. Eqns (\ref{rho_c} - \ref{prob2}) have been expressed in 
terms of the fidelity $F$ of the $|\Phi_-\rangle$ state generated, 
the attenuation parameters $l/l_0$ and the detection efficiency rather than $\beta$. The initial displacement $\beta$ can be 
expressed in terms of $F$, $l/l_0$, $P$ and $\eta^2$. We also need to point out that Eqn (\ref{rho_c}) is a mixture of only two Bell states, 
$|\Phi_\pm\rangle$, which has important advantages in entanglement purification. Such states are much more efficient to purify.

\begin{figure}[!htb]
\begin{center}\includegraphics[scale=0.8]{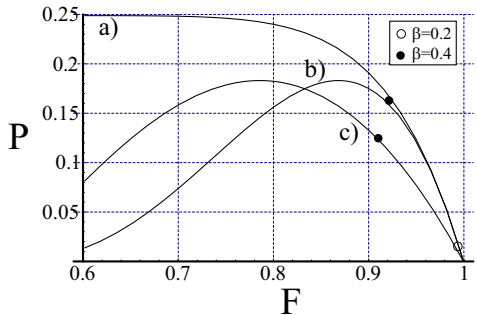}\end{center}
\caption{Plot of the probability $P[F,l/l_0]$ of successfully establishing an entangled pair at an attenuation 
length of $l/l_0=0.8$ (20 km in commercial fiber) versus fidelity for the different measurement strategies: a) odd state cat 
projector CSP, b) ideal SPD and c) SPD with $\eta^2=0.9$.  
}
\label{fig-fid-prob}
\end{figure}

In Fig (\ref{fig-fid-prob}), we plot these probabilities $P[F,l/l_0]$ versus fidelity for both measurement 
strategies (with $\eta^2=0.9,1$) for an attenuation length $l/l_0=0.8$. Our idealized cat projector (CSP) 
allows a nice range of fidelities to be achieved, ranging from $F=1/2$ to $F=1$.  For higher 
fidelities, lower success probabilities are achieved. The major issue with the cat projector is that it is difficult to 
implement in practice, but for the moderate to high fidelity regimes, the single photon detection scheme 
closely follows the cat projector results and so is an excellent compromise. It also highlights
that single photons (and hence single-photon sources) are not needed for creating high fidelity distributed Bell pairs. The sending of 
coherent states (whether weak or strong) can achieve the same goal. High fidelity pairs can be generated over long distances 
but at the expense of the success probability. 
Other measurements techniques (such as homodyne, bucket or vacuum detection) also result in entangled states, but these states 
tend be composed of more than two Bell states, and so are more difficult to purify. However, they can provide 
success probabilities greater than one half but result in low fidelity
final states.

We now have a highly tunable segment where one can dynamically vary the quality of the base-level entangled pair from low to high 
fidelity utilizing non-unit efficiency single-photon detection. These pairs can be used in a repeater protocol to create long-distance
entangled pairs. The basic repeater protocol works as follows:
multiple copies of lower fidelity base-level pairs are purified 
to create high fidelity pairs, then entanglement swapping of the high fidelity pairs between adjacent repeater nodes 
creates longer distance but lower fidelity pairs. These resulting pairs  can then purified to high fidelity pairs and entanglement swapping gives 
even longer range pairs. The procedure is iterated until pairs over the desired length are obtained. The purification and entanglement 
swapping protocols require efficient local two-qubit C-Z (or CNOT) gates. In our architecture these can be achieved using 
another sequence of controlled displacement operations  $D(i\beta_2 \sigma_{z2}) D(\beta_1 \sigma_{z1}) D(-i\beta_2 \sigma_{z2}) 
D(-\beta_1 \sigma_{z1})\equiv \exp \left[2 i\; \beta_1\beta_2\;\sigma_{z1} \otimes \sigma_{z2}\right]$  
with $\beta_1$ and $\beta_2$ satisfying $\beta_1 \beta_2=\pi/8$ \cite{spiller06,loock07}. These quantum bus (qubus) 
based local gates are needed throughout the protocol, and so the more efficient and robust they are the better our 
overall performance.

One of the major issues for performance becomes the chosen quality of base-level entangled pairs for our 
lowest-level segment. Conventional quantum repeater wisdom generally suggests that before one performs entanglement
swapping to create longer pairs, one should purify the noisy base-level pairs as best as possible. For our discussions
here we will set a working fidelity of $F=0.98$ before attempting entanglement swapping. There are now a 
number of ways we can use our tunable segment to achieve this required fidelity. We could just directly create a 
pair of that fidelity (see Table \ref{table-1}), or we could create lower-fidelity pairs and purify them using 
standard protocols\cite{dur07,bennett93,dur99,pan01}. Which approach is best will depend on 
both the probability of generating the entangled pairs $P_g$ and the purification probability $P_{pur}$. For a 
one-round purification protocol with no limitation on the physical resources, the effective probability of 
generating the final fidelity pair (per channel) is given by $P_{eff}=P_g P_{pur}/2$.  The purification probability 
depends critically on the form of the initial entangled pairs, and so our engineering of Eqn (\ref{rho_c}) above is 
very advantageous. A mixture of two Bell states is much easier to purify than more general mixed states. In this situation, 
two copies of $\rho (F)$ can be purified to a new entangled $\rho (F'=F^2/(1-2 F+2F^2))$ 
with a $P_{pur}(F)=F^2+(1-F)^2$\cite{pan01}.  This gives an overall probability of success for generating a $F=0.98$ 
pair from two $F=0.9$ pairs  of $P_{eff}=0.0547$. In comparison, we have generation probabilities of $0.03374$ for 
directly manufacturing the $F=0.98$ pair. Thus, using lower fidelity pairs and purification improves our overall probability of 
generating the final pair, assuming efficient local gates. 

\begin{table}[!htb]
\begin{tabular}{|c||c|c|c|}
	\hline
	&  F=0.98    &   F=0.9  & F=0.75   \\
	\hline \hline
$P_{SPD}[F,0.4,0.9]$   & 0.05076 & 0.16693 & 0.13996 \\
$P_{SPD}[F,0.8,0.9]$   & 0.03374 & 0.13338 & 0.17970 \\
$P_{SPD}[F,1.6,0.9]$   & 0.01494 & 0.07157 & 0.15528 \\
	\hline
\end{tabular}
\caption{Success probability of generating a Bell state of fidelity $F$ conditioned on 
single photon detection ($\eta^2=0.9$).}
\label{table-1}
\end{table}

Does a multiple-round purification protocol improve our results?
Moving to multiple rounds, there are a number of choices for how to
implement the protocol\cite{dur07,dur99}.
As an illustration, consider a symmetric purification protocol\cite{dur99}
where four F = 0.75 pairs are purified to the F = 0.98 pair. In this
case, we have an effective generation probability of Pg = 0.0236 which
is lower than the probability achieved by the single-round
protocol. The difference is primarily due to the lower probability of
successfully purifying lower fidelity pairs.

We now need to turn our attention to a more detailed discussion of the
physical resources required for creating our long distance pairs. It
is important to consider both the spatial and temporal resources
necessary. The tunability of our source allows us significant
flexibility in how we use resources. We can implement the minimum
physical resource strategies (two qubits per station) developed by
Harvard\cite{childress06}, as well as the modest physical resource
{sixteen qubit per station} schemes of van Loock {\it et
  al.}\cite{loock06,ladd06}. Due to our low to moderate probabilities
of successfully creating the base pairs between stations, the minimum
physical resource approach will have significant performance issues
due to this distribution bottleneck. Significant time will be spent
idle waiting for the base pairs to be successfully created between all
stations. However, by allowing moderate physical resources, we can
simultaneously attempt to create multiple base pairs per segment, and
so reduce the time waiting for the necessary resources to become
available. This approach will dramatically increase the throughput of
the entire repeater chain. To quantify this degree of the improvement,
we have performed a Monte Carlo simulation of a nested entanglement
protocol over 51.2 $l/l_0$ (1280 km) for varying numbers of qubit per
station with dynamical resource allocation\cite{collins07}. The 
results are presented in Table(\ref{table-2}).  First,
they show that, of the three segment lengths considered, the best
results were obtained from the 0.8 $l/l_0$ (20km) situation.  For
longer distances, the initial success probability drops dramatically,
and for shorter segment lengths, errors in local gates have a
significant impact.  We also found that the protocol could run with
local gate error rates exceeding 1\%, though in this situation the
generation rate falls to only a few pairs per second. Raising the
number of qubits in each half station also gives a slight improvement
compared to stacking smaller repeater nodes in parallel. Still, our
results show a good generation rate with 8-16 qubits per half node.
In the scheme of van Loock, with 16 qubits per half node with 0.4
$l/l_0$ segment spacing (4096 total qubits), a rate of 15 F=0.98 Bell
pairs/second was achieved.  Using our new scheme (for the same total
number of qubits), we achieve a rate of 3190 pairs/second, an
improvement of over two orders of magnitude. Our new scheme also 
produces relatively high throughput of 437 pairs at a link distance of 
40km compared with zero for the orginal van Loock case. In the new scheme, 
the fidelity remains high over long distances, but the probability 
of success declines. Finally, some of our improvements in the protocol 
are obtained by tuning the base-level fidelity to optimize the number 
of purification rounds before entanglement swapping.

\begin{table}[!htb]
\begin{tabular}{|c||c|c|c|}
	\hline
qubits	&  0.4 $l/l_0$  &  0.8 $l/l_0$  & 1.6 $l/l_0$  \\
half station &  &   &  \\
	\hline \hline
8     & 520 (2048) & 693 (1024) & 193 (512) \\
16    & 1097 (4096)&  1528 (2048) & 437 (1024)) \\
32    & 2297 (8192)&  3190 (4096) & 987 (2048)) \\
	\hline \hline
\end{tabular}
\caption{Rate of final generation of distant 51.2 $l/l_0$ (1280 km) entangled pairs 
of minimum fidelity $F=0.98$ resulting from a nested entanglement protocol with 8, 16 and 32  
qubits per half-station. The stations are separated by either 0.4 $l/l_0$, 0.8 $l/l_0$ 
or  1.6 $l/l_0$ attenuation lengths.The number in the brackets indicate the total 
numbers of qubits over all repeater stations. We have assumed an initial fidelity reduction in 
the base pairs due to loss in the fiber (assumed to be 0.17dB/km) and distortion of 0.1\%  due 
to local losses during the measurement-free displacement-based C-Z gate.}
\label{table-2}
\end{table} 

{\it To summarize} we have shown how to implement a high bandwidth
quantum repeater using the fundamental atom-light interactions in
quantum optics, through a qubus-mediated entangling operation. This
new approach refutes the criticism that hybrid repeaters cannot adapt
to a wide range of losses and offers flexibility on fidelity and
entanglement success probability.  In this hybrid scheme, the only
required interactions are controlled displacement operations between
the light field and qubit.  These controlled displacement operations
result in the distribution of moderate- to high-fidelity entanglement
between the qubits in the repeater stations, conditioned on
single-photon detection of the qubus mode.  They also can implement a
deterministic C-Z gate for use in all the purification and
entanglement swapping steps.  Such tools allow for the natural design
of a scalable and homogeneous quantum repeater network and thus a
distributed computation with many concurrent steps happening in
different locations. Using high-efficiency single photon detection, we
have shown that long-distance communication rates over 1000
pairs/second with a fidelity above 98\% are possible with only modest
resources.

\noindent {\em Acknowledgments}: We thank T. Ladd for the use of the 
base simulation code upon which our simulations are based and also thank 
him, L. Jiang, P. van Loock, T. P. Spiller and J. M. Taylor for valuable 
discussions. This work was supported in part by MEXT and NICT in Japan 
and the EU project QAP.


\begin{thebibliography}{99}
\bibitem{footnote} The sequence of operations $D(\alpha\cos\theta)e^{-i \theta \sigma_z
a^{\dagger} a} D(-2\alpha)$ $e^{i \theta \sigma_z a^{\dagger}a}D(\alpha\cos\theta)$ generates 
a controlled displacement of the form $D\left(2i \alpha \,\sin\theta\,\sigma_z\right)$. Here 
the controlled rotations operations $e^{-i \theta \sigma_z a^{\dagger} a}$ are achieved by 
operating in the dispersive limit while $D(\alpha)$ are unconditional displacements.
\end{thebibliography}
\end{document}